\documentclass[preprint,showpacs,preprintnumbers,amsmath,amssymb,prb]{revtex4}
\usepackage{graphicx}
\usepackage{dcolumn}
\usepackage{bm}
\usepackage{epsfig}
\begin{document}

\preprint{PREPRINT}

\title{Entropy, diffusivity and the energy landscape of a water-like fluid}

\author{ Alan Barros de Oliveira}
\affiliation{Departamento de F\'{i}sica, Universidade Federal de Ouro Preto,
Ouro Preto, MG, 35400-000, Brazil} 
\email{oliveira@iceb.ufop.br}

\author{Evy  Salcedo}
\affiliation{ Instituto de F\'{\i}sica,
Universidade Federal do Rio Grande do Sul, Porto Alegre, RS, 
91501-970, BRAZIL}
\email{esalcedo@if.ufrgs.br}

\author{Charusita Chakravarty}
\affiliation{Department of Chemistry,
Indian Institute of Technology-Delhi,
New Delhi, 110016, India}
\email{charus@chemistry.iitd.ernet.in }

\author{ Marcia C. Barbosa}
\affiliation{ Instituto de F\'{\i}sica,Universidade Federal do Rio Grande do Sul,  Porto Alegre, RS, 
91501-970, Brazil} 
\email{marcia.barbosa@ufrgs.br}
\begin{abstract}

Molecular dynamics simulations and instantaneous normal mode (INM)
analysis of a fluid with  core-softened pair interactions
and  water-like liquid-state anomalies are performed to obtain
an understanding of the relationship between thermodynamics, transport
properties and the potential energy landscape. Rosenfeld-scaling
of diffusivities with the thermodynamic excess and pair correlation entropy
is demonstrated for this model. The INM spectra are shown to carry information about
the dynamical consequences of the interplay between length
scales characteristic of anomalous fluids, such as bimodality of
the real and imaginary branches of the frequency distribution. 
The INM spectral information is used to partition the liquid
entropy into two contributions associated with the real and imaginary
frequency modes; only the entropy contribution from the imaginary branch
captures the non-monotonic behaviour of the excess entropy and diffusivity
in the anomalous regime of the fluid.
\date{\today}

\end{abstract}


\maketitle

\section{Introduction}
The potential energy surface (PES) , $U({\bf r})$,  is  the configurational 
energy of a system of $N$ particles as a function of the 
$3N$-dimensional position vector, ${\bf r}$. Energy landscape approaches focus
on the connections between crucial topographical features of the PES
and the thermodynamic and kinetic properties of liquids.\cite{sw82,ds01,djw03} Such studies have
focused largely on simple liquids where interactions are  dominated by 
strong, short-range, repulsions with weak, long-range attractions.\cite{hm02} 
Since the hard-sphere fluid with a single length scale is a very good
zeroth-order model for such systems,  the relationship between the energy
landscape, thermodynamics and mobility is expected to be relatively simple
compared to anomalous fluids. 
Recent work, however, demonstrates that softening  the core-repulsions
in liquids with isotropic, pair-additive interactions allows one to 
generate a range of anomalous behaviour that mimics the behaviour of 
structurally more complex fluids. \cite{Ja98,Wi06,Xu05,Ol07,Ol06a,Ol06b,Fo08,Eg08, Mi06b,Er01,Ne01,Ne02b,Zh08,Ba09,Ca03}

In this study, we explore the energy landscape of
such  a core-softened fluid to understand the  microscopic origins of 
water-like liquid state anomalies. Our analysis of the potential 
energy surface focuses on understanding
the  entropy scaling relationships that are
very useful for connecting  structure, mobility and entropy 
for a wide range of simple 
\cite{yr771,yr772,yr99,md96,has00,sag01,sag05} and
anomalous
\cite{scc06,ac09,agc09,etm06} liquids, 
 confined fluids  and polymeric melts.\cite{gpmc08}
The excess entropy ($S_{ex}$) measures the reduction in the entropy 
($S$)
of a liquid relative to an ideal gas ($S_{id}$) at the same temperature and
density due to structural correlations. The effect of fluid structure on 
the entropy can be formally expressed  as:
\begin{equation}
S_{ex}=S-S_{id}=S_2 +S_3 + \dots
\end{equation}
where $S_n$ is the entropy contribution due to $n$-particle spatial 
correlations.\cite{hsg52,ng58,hjr71,dcw87,be89} The pair correlation contribution  to
the excess entropy per particle
 of a one-component fluid of structureless particles is 
given by
\begin{equation}
s_2^* = -2\pi\rho \int_0^\infty 
\{g(r)\ln g(r) -\left[ g(r)-1\right] \} r^2 dr 
\label{s2}
\end{equation}
where $g(r)$ is the radial distribution function and $s_2^*=S_2/Nk_B$. The 
structural correlations
which lower the entropy may intuitively be expected to reduce mobility
by enhancing cage effects due to formation of shells of neighbouring particles.
This correlated decrease  in entropy and  mobility 
can be semi-quantitatively captured
through excess entropy scaling relations of the form
\begin{equation} X^*=A\exp (\alpha s_{ex}^*)\; ,
\end{equation}
where $X^*$ are dimensionless transport properties with either macroscopic
(Rosenfeld) or microscopic (Dzugutov) reduction
parameters,  $s_{ex}^*$ is the excess entropy per particle in units of $k_B$.
The scaling parameters, $\alpha$ and $A$, depend on the functional form 
of the underlying interactions.\cite{yr771,yr772,yr99,md96,has00,sag01,sag05}
In the case of simple liquids, the excess entropy scaling parameters
can be approximately set as $A\approx 0.6$ and $\alpha
\approx 0.8$. In addition, for such fluids, the pair correlation entropy per
particle, $s_2^*$,
typically represents 85\% to 90\% of the total excess entropy.

This paper focuses on the connection
between the potential energy landscape (PEL) of a fluid and  Rosenfeld 
excess entropy 
scaling of transport properties.  Liquids in the stable, as opposed
to the strongly supercooled regime, are characterized by 
a very high degree of   connectivity between basins of local minima.
This implies that the diffusivity, corresponding to the probability that 
a particle will make a successful move from its current position, will  be
proportional to   the number of accessible configurational states, or
$\exp(\alpha s_{ex}^*)$. In order to develop a quantitative test of this
intuitive picture, the diffusivity and/or the entropy must be correlated with
landscape-based quantities that are 
sensitive to basin connectivity. We also require an energy landscape
approach that does not presume a time-scale separation between intra- and
inter-basin motions. An existing energy landscape approach that is simple
to implement and satisfies these requirements is the Instantaneous Normal Mode 
(INM)  approach.\cite{sm96,rms95,tk97} In the INM approach, the key quantity
is the  ensemble-averaged curvature distribution of the PES sampled by the 
system.  For a system of $N$ particles, the mass-weighted Hessian associated with 
each instantaneous configuration is diagonalized to yield 3N normal mode
eigenvalues and eigenvectors and the ensemble-average of this
distribution is referred to as the INM spectrum.  The short-time 
dynamics of the liquid can be derived from the INM spectra. 
Unlike in a crystalline solid, the INM spectrum
of a liquid will have a substantial fraction of unstable modes with 
negative eigenvalues, corresponding to inter-basin crossing modes or shoulder
regions within the same inherent structure basin.
 The diffusivity is strongly correlated with
the properties of the INM spectrum, specially the fraction of imaginary
frequencies, in both simple liquids, such as Lennard-Jones 
and Morse,\cite{sc01a,sc021,sc022} as well as molecular liquids, such 
as CS$_2$ and H$_2$O.\cite{cfsos94,lssss00,rm97} A refinement
of the INM approach, shows that  interbasin crossing or double-well
modes are critical for diffusional motion.\cite{lks98}

Here we study the instantaneous normal mode spectra of a 
liquid bound by isotropic, core-softened pair interactions which shows 
water-like structural, density, entropy and diffusional anomalies.
The  thermodynamic and transport properties of such a liquid
is representative of structurally more complex anomalous liquids, 
including  water \cite{Er01,Ne01,Ne02b,Mu05} and
other tetrahedral liquids, such as
 Te, \cite{Th76} Ga, Bi,~\cite{LosAlamos} S,~\cite{Sa67,Ke83}
Ge$_{15}$Te$_{85}$,~\cite{Ts91}
silica,~\cite{An00,Ru06b,Sh02,Po97}
silicon,~\cite{Ta02} and BeF$_2$.~\cite{An00,scc06,asc07,ac07,ac09,agc09}
Section II describes our
 core-softened model fluid  with 
isotropic interactions consisting of a sum of 
Lennard-Jones and
Gaussian terms. The continuous nature of the pair interaction makes it
very convenient for energy landscape analysis. The liquid state anomalies
of this model, which have been described in detail elsewhere,\cite{Ol06a,Ol06b} are summarized. In Section III, we provide
a summary of the relevant features of instantaneous normal mode analysis.
We also address the  
possibility of extracting thermodynamic quantities, including
excess entropy, from INM spectra which has so far not been discussed in
the literature.
Section IV presents our results and
Section V contains the conclusions.

\section{The model} 
\label{sec:model}

\subsection{Potential energy surface}
We consider a three-dimensional, core-softened fluid with isotropic
pair interactions given by:
\begin{equation}
U(r)=4\epsilon\left[\left(\frac{\sigma}{r}\right)^{12}-
\left(\frac{\sigma}{r}\right)^{6}\right]+
a\epsilon
\exp\left[-\frac{1}{c^{2}}\left(\frac{r-r_{0}}{\sigma}
\right)^{2}\right].
\label{eq:potential}
\end{equation}
Equation (\ref{eq:potential}) shows that the pair interaction
is composed of a Lennard Jones  term, with
characteristic energy and length scale parameters corresponding to 
$\epsilon$ and $\sigma$ respectively, plus a Gaussian well centered at a
pair separation $r_{0}$ with depth $a\epsilon$ and width $c\sigma$. 
In this work we use the parameters for  Eq.~ (\ref{eq:potential}) 
as $a=5,$ $r_{0}/\sigma=0.7$ and $c=1$. This set of parameters generates 
a core-softened potential  with a very small attractive minimum at 
$r\approx 3.8$ and  a soft, repulsive core lying between 
$\sigma$ and 3$\sigma$. 
All quantities in this paper are reported in reduced units with
the $\sigma$ and $\epsilon$ as the reduced units of length and energy 
respectively.
The water-like structural, density and diffusional anomalies
of this model are briefly described in this section in order to provide
a background to the INM results presented in Section IV.

\subsection{Molecular Dynamics Simulations}

Classical molecular dynamics (MD) simulations were used to study the 
model fluid
described in the previous subsection. 
$N=500$ identical, structureless particles of mass $m$ were confined in a 
cubic box, of volume $V$, with periodic boundary conditions
in all directions. All MD simulations were performed in the canonical
(NVT) ensemble with a time step of $0.002\sigma \sqrt{m/ \epsilon}$. 
A Nos\'e-Hoover thermostat  with the coupling parameter equals to 2 was used 
to maintain the temperature.  All simulations were initialized with the system
in a face centered cubic configuration and further equilibrated over 
250 000 steps for each 
temperature, T, and density, $\rho=N/V$.  After the equilibration period was over,
additional 500 000 steps were used to sample the system.  A cutoff radius 
$r_c = 3.5 \sigma$ was employed for the  potential Eq. (\ref{eq:potential}).
Diffusivities were computed using the Einstein relation.
At each state point, 100 configurations were sampled and used to
construct the instantaneous normal mode spectra and associated quantities.
We repeated the calculation for some state points using 500 configurations 
and found no significant difference. 

\subsection{Density, Diffusional and Structural Anomalies}

\begin{figure}[htbp]
   \centering
   \includegraphics[clip=true,scale=0.5]{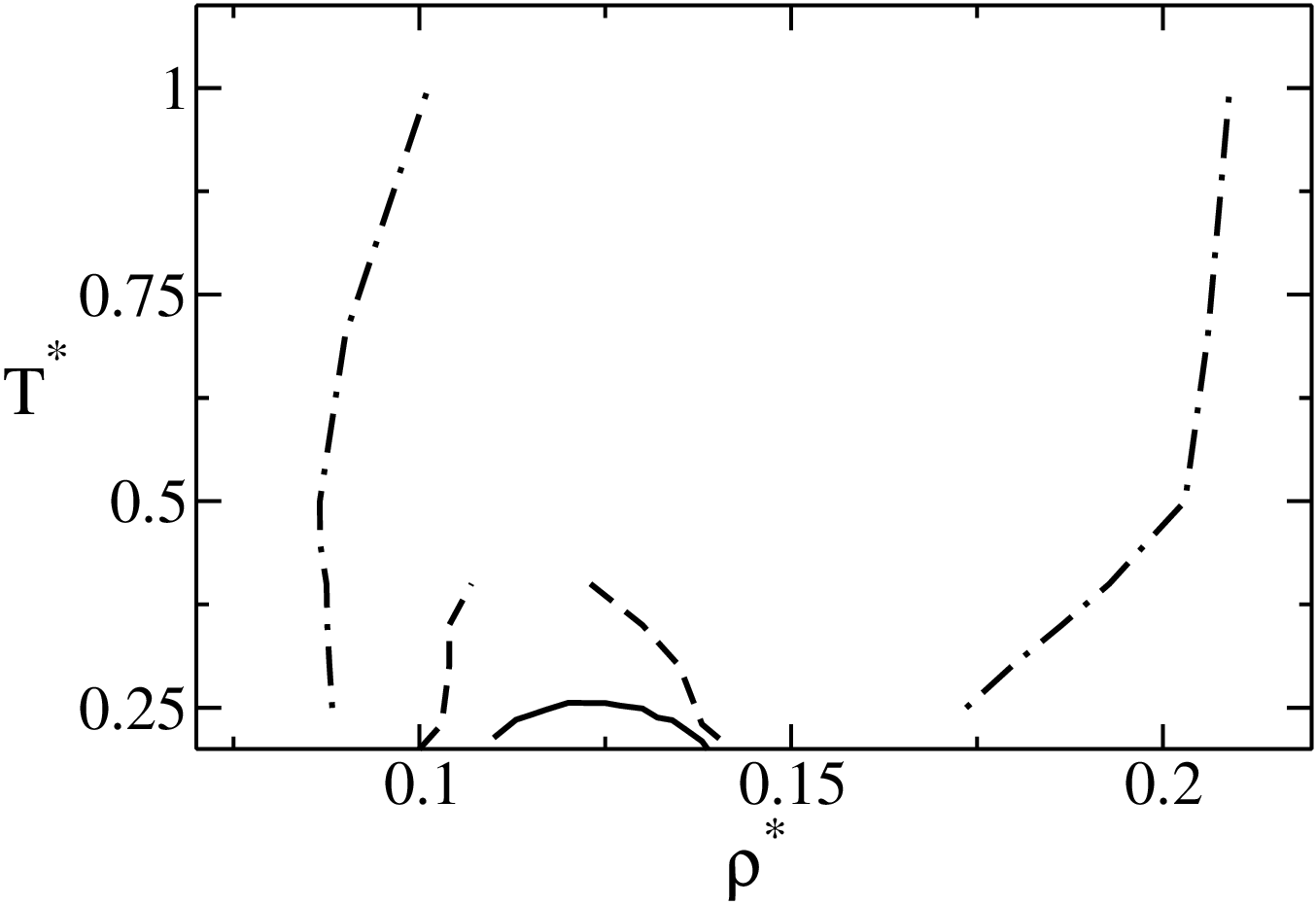} 
   \caption{Density versus temperature phase diagram for the 
model studied. The solid line limits the region of 
density anomaly, the dashed lines illustrate the region
of diffusion anomaly and the dot-dashed lines 
show the region of structural anomaly.}
   \label{fig:phase-diagram}
\end{figure}

Figure~\ref{fig:phase-diagram} illustrates  the regions associated
with the density, diffusional and structural anomalies of the model
fluid studied here  in  the density-temperature planes.  The region of
density anomaly corresponds to state points for which 
$(\partial \rho /\partial T)_P > 0$ and is bounded by the locus of
points for which the thermal expansion coefficient is zero.
The  translational diffusion coefficient
 as a function of $\rho^*=\rho \sigma^3$ goes as follows.
For the low temperature isotherms, the  diffusivity increases as the density 
is lowered, reaches a maximum at $\rho_{D{\rm max}}$ 
and decreases
until it reaches a minimum at $\rho_{D{\rm min}}$.
The locus of extrema in the $D(\rho )$ curve mark the boundaries of the
region of diffusional anomaly, as shown in Figure~\ref{fig:phase-diagram} 
using dashed lines.

The region of structural anomaly of core-softened fluids is defined most 
simply 
using the translational or pair correlation order metric, defined as \cite{Er01}
\begin{equation}
t\equiv\int_{0}^{\xi_{c}}|g(\xi)-1|d\xi,
\label{eq:trans}
\end{equation}
\noindent where $\xi\equiv r\rho^{1/3}$ is the 
interparticle separation scaled by the mean interparticle distance,
$g(\xi)$ is the radial distribution function
and $\xi_c$ is a scaled  cut-off distance. 
In this work, we use  $\xi_c=\rho^{1/3} L/2,$
where $L=V^{1/3}.$  For a completely uncorrelated system (ideal gas) $g=1$
and $t$ vanishes. In a crystal, the presence of  long-range translational 
($g\ne1$) implies that $t$ depends on choice of $\xi_c$.
In simple liquids, $t$ increases with isothermal compression. In  liquids
with water-like anomalies at low temperatures, $t$ shows a non-monotonic
behaviour. At a given temperature $T$, a structurally anomalous regime can
be defined between densities $\rho_{t-max}(T)$ and  $\rho_{t-min}(T)$ 
corresponding to locations of the maxima and minima in the translational 
order.
In this structurally anomalous  regime, shown using dot-dashed lines in
Fig.~\ref{fig:phase-diagram}, an increase in density induces a 
decrease in translational order. The nested structures of the anomalous
regions is evident with  the structurally anomalous regime enclosing
the diffusion anomalous region which in turn encloses the region of
density anomaly.

\section{Instantaneous Normal Modes Analysis}

In this section, we define the instantaneous normal mode spectrum and
explore how it can be used to connect the energy landscape of a liquid
with its thermodynamic entropy. The 
potential energy of  configuration ${\bf r}$ near ${\bf r_0}$ 
can be written as a Taylor expansion of the form:
\begin{eqnarray}
U({\bf r})=U({\bf r_0})-{\bf F} \bullet {\bf z}+\frac{1}{2} {\bf r^T} 
\bullet {\bf H} \bullet {\bf z}
\label{eq:expansion}
\end{eqnarray}
where ${\bf z_i}=\sqrt{m_i}({\bf r_i}-{\bf r_0})$ are the mass-scaled position coordinates of a particle $i$.
The first and second derivatives of $U({\bf r})$ with respect to the vector 
${\bf z}$  are the force and the Hessian matrix, denoted by 
${\bf F}$ and ${\bf H}$ respectively.
The eigenvalues of the Hessian ${\bf H}$ are $(\{ \omega_i^2\}, i=1,3N)$
representing  the squares of normal mode frequencies, and ${\bf W}({\bf r})$ 
are the corresponding eigenvectors. In a stable solid,  ${\bf r_0}$ can be
conveniently taken as the global  minimum of the potential energy
surface $U(R)$, which implies that ${\bf F}=0$ and ${\bf H}$ has
only positive eigenvalues corresponding to oscillatory modes.  The INM approach
for liquids  interprets ${\bf r}$ as the configuration at time $t$
relative to the configuration ${\bf r_0}$ at time $t_0$. Since typical
configurations, ${\bf r_0}$ are extremely  unlikely to be local minima,
therefore ${\bf F}\neq 0$ and
${\bf H}$ will have negative eigenvalues. The negative eigenvalue modes are
those which sample negative curvature regions of the PES,
including barrier crossing modes.
The ensemble-averaged INM spectrum, $\langle f(\omega) \rangle$,  is 
defined as
\begin{eqnarray}
f(\omega)=\langle \frac{1}{3N} \sum_{i=1}^{3N}
\delta(\omega-\omega_i)\rangle.
\label{eq:rho}
\end{eqnarray}
Quantities that are convenient for characterizing the instantaneous normal
mode spectrum are: (i) the fraction of imaginary frequencies, namely
\begin{eqnarray}
F_{im}=\int_{im}^{} f({\omega}) d\omega
\label{eq:Fi}
\end{eqnarray}
where the subscript $in$ means that the integral is
performed only in the imaginary branch; (ii) the
fraction of real frequencies, that is
\begin{eqnarray}
F_{r}=\int_{r}^{} f({\omega}) d\omega
\label{eq:Freal}
\end{eqnarray}
where the subscript $r$ indicates that
the integral is performed only in
the real branch and (ii) the mean square or Einstein 
frequency, $\omega_E$,  given by
\begin{eqnarray}
\omega_E^2&=&\int_{}^{}\omega^2  f(\omega) d\omega \nonumber \\
&=& \frac{\langle Tr {\bf H} \rangle}{m(3N-3)} 
\label{eq:omegaE}
\end{eqnarray}
where the last equality comes from using Eq.~(\ref{eq:rho}) and
$\langle Tr {\bf H} \rangle $ is the ensemble-averaged value of the trace
of the Hessian. 

The simplest approximation to the entropy that can be derived from the
INM approach is to consider a liquid as a collection of 3N simple harmonic 
oscillators vibrating at the Einstein frequency. The entropy of a
one-dimensional harmonic oscillator with frequency $\omega$ is given by:
\begin{equation}	
s_{\omega}/k_B= 1-\ln ({\beta \hbar\omega)}.
\label{eq:s-omega}
\end{equation}
The entropy of an ideal gas of N particles in 3-dimensions will be
given by:
\begin{equation}
\frac{S_{id}}{Nk_B}=\frac{5}{2}-\ln(\rho\Lambda^3)\; ,
\label{eq:s-ideal}
\end{equation}
where $\Lambda =h/\sqrt{2\pi mk_BT}$ is the thermal de Broglie wavelength.
In three dimensions the entropy of the harmonic oscillators given
by Eq.~(\ref{eq:s-omega}) is multiplied by 3. The  entropy per
particle of the
harmonic oscillator within Einstein approximations becomes $s_{\omega_E}$.
In this case the excess entropy of the 
Einstein model of the  liquid  is  given 
by the subtraction of the ideal gas entropic 
contribution, Eq.~(\ref{eq:s-ideal}), from $Ns_{\omega_E}/k_B$  to give
\begin{equation}
S_{\mathrm{ex}}^{\omega_E}/N k_B = \frac{1}{2} 
+ \frac{3}{2} \ln \left( \frac{2 \pi k_B T \rho^{2/3}}{m \omega_{E}^2} \right)\; ,
\label{eq:s-e}
\end{equation}
where $\omega_E$ is given by Eq.~(\ref{eq:omegaE}).
This expression for the excess entropy forms the basis of quasi-harmonic cell
model approaches to understand entropy-scaling of transport properties 
which have had only limited success.\cite{yr99} In this study, 
we compare the Einstein frequency-based expression for the excess entropy 
with the pair correlation
entropy to obtain a better microscopic insight into the differences.

An alternative approach is to consider the liquid to be composed
on average of a set of $3NF_r$ harmonic oscillators and  a set of
$3NF_{im}$ degrees of freedom associated with the imaginary or unstable modes.
The total thermodynamic entropy of the liquid  can be 
written as a sum of
contributions from the real and imaginary branches:
\begin{equation} 
S=S_{r} + S_{im}.
\label{eq:s}
\end{equation}

Using equation(\ref{eq:s-omega}), 
$S_{r}$ can be obtained by
integrating the real branch of the INM distribution as follows: 
\begin{equation}
S_{r}/k_B = 3N\int_{r} 
f(\omega) s_{\omega}(\omega) d\omega, 
\label{eq:sreal}
\end{equation}
\noindent where $f(\omega)$ is the  INM probability 
density at frequency
$\omega$ given by Eq.~(\ref{eq:rho}). Contribution of
the imaginary modes to the entropy must then be
\begin{equation}
S_{im}=S - S_r = S_{id}+S_{ex}-S_{r}\; .
\label{eq:simag}
\end{equation}

In the present work, we use the above equations to estimate $S_{r}$ and
$S_{im}$; the per particle values of these quantities in units of $k_B$ are labeled
$s_r^*$ and $s_{im}^*$.  To our knowledge, this decomposition of the entropy has not 
been used previously. A somewhat parallel approach was used by Goddard\emph{ et al.}
to treat the entropy of a liquid as a sum of contributions from a harmonic
component and a hard-sphere fluid component.\cite{goddard}

\section{Results}

\subsection{Excess Entropy and Diffusivity}

\begin{figure}[htbp]   
\centering   
\includegraphics[clip=true,scale=0.5]{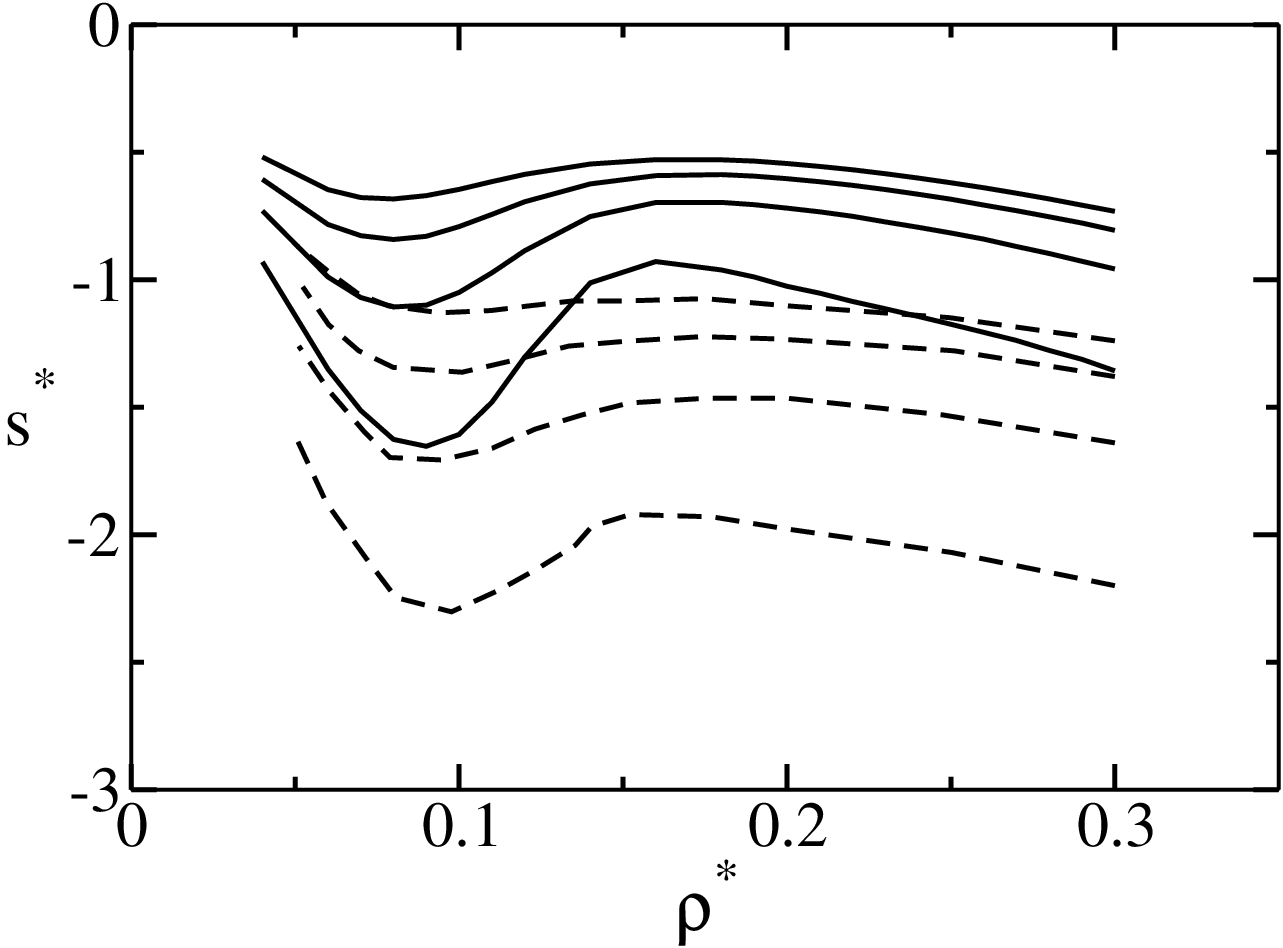}
\caption{The pair correlation entropy, $s_2^*$ (solid lines), and the excess entropy,
$s_{ex}^*$\cite{etm06}, (dashed lines)
against density for fixed 
temperatures for $T^*=0.2,0.3,0.4,0.5$ from bottom to top. }
\label{fig:s2-sex}
\end{figure}

The
 excess entropy is defined as the difference between the entropy
of the real fluid and that of the
ideal gas at the same temperature and density. 
Figure~\ref{fig:s2-sex} illustrates the density dependence of
the excess  (represented in the figure by dashed lines)  and pair correlation entropy (represented in the figure by solid lines) for four different
isotherms. The values of the thermodynamic excess entropy, $s_{ex}^*$,
have been taken from Mittal \emph{ et al.} work.\cite{etm06}

The $s_{ex}^*(\rho )$ curves at low temperatures show
a pronounced excess entropy anomaly, corresponding to 
a rise in excess entropy on isothermal compression.  Such an entropy
anomaly is characteristic of water-like 
liquids \cite{scc06,ac09,agc09,Mi06b} and
contrasts with the behaviour of simple liquids where free volume arguments
are sufficient to justify a monotonic decrease in entropy on
isothermal compression. Figure 2 also compares $s_{ex}^*(\rho )$ and
$s_2^*(\rho )$ curves at four temperatures.
It is evident that $s_2^*$ essentially captures
the anomalous behavior present in $s_{ex}^*$. The effect
of the higher-order multiparticle correlations terms in $s_{ex}^*$
is  to generate a downward shift in the values of the entropy and
to attenuate the entropy anomaly. In the case of simple liquids, the
residual multiparticle entropy 
(RMPE), $\Delta s^* =s_{ex}^*-s_2^*$, is typically
of the order of 10-15\% of $s_{ex}^*$ for a fairly wide range of
densities. Clearly in the case of the core-softened modeled fluids,
the residual multiparticle entropy
 contribution is larger in magnitude and more strongly density-
dependent. The anomalous pair entropy regime 
at a given temperature is an  interval of densities 
$\rho_{s2_{max}}<\rho<\rho_{s2_{min}}$ within which 
$(\partial S_{2}/\partial\rho )_T > 0$. This can be  identified
from the locus of extrema in $s_2(\rho )$ shown in Fig.~\ref{fig:extrema}.

\begin{figure}[htbp]   
\centering   
\includegraphics[clip=true,scale=0.35]{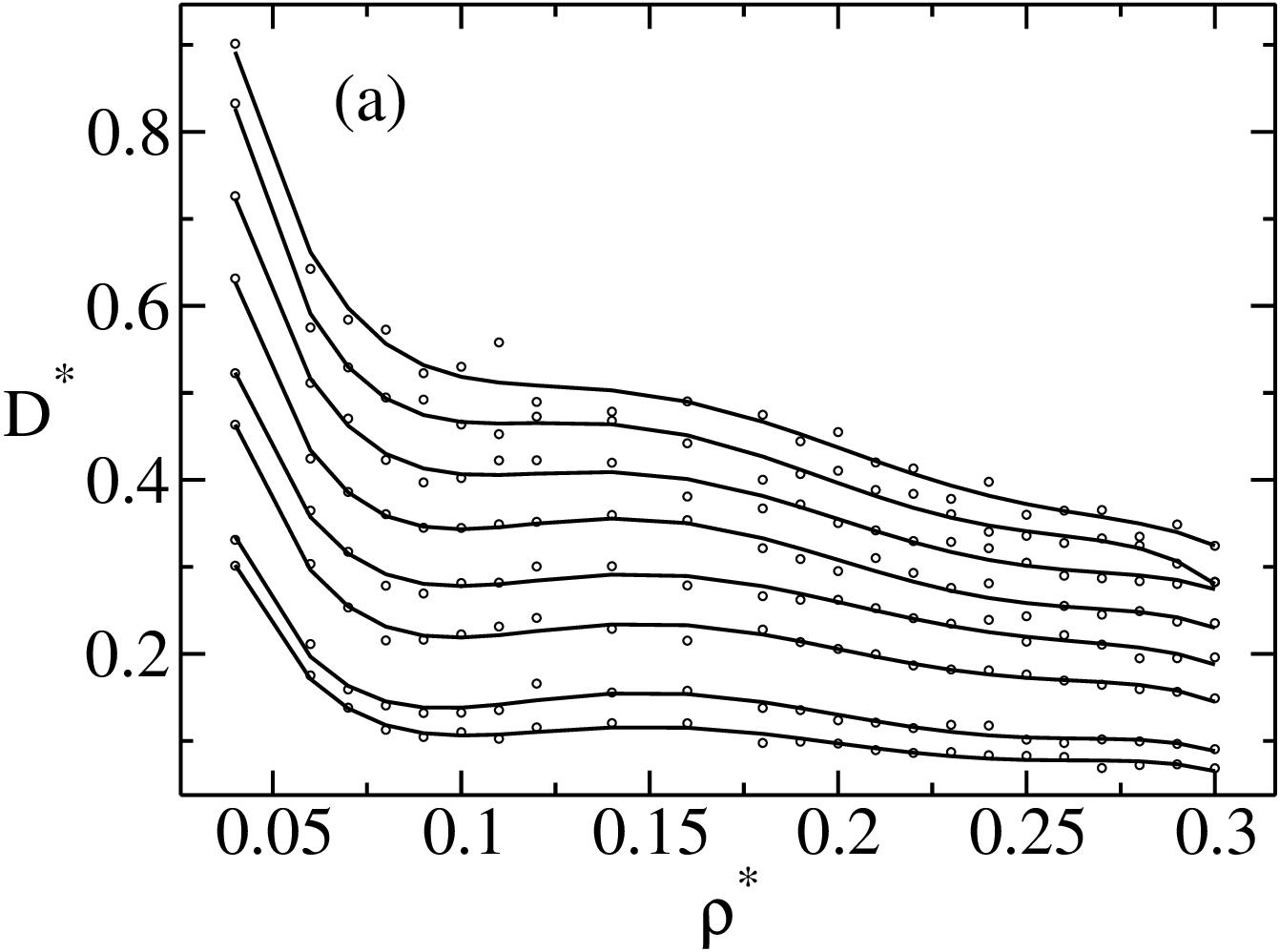}\includegraphics[clip=true,scale=0.35]{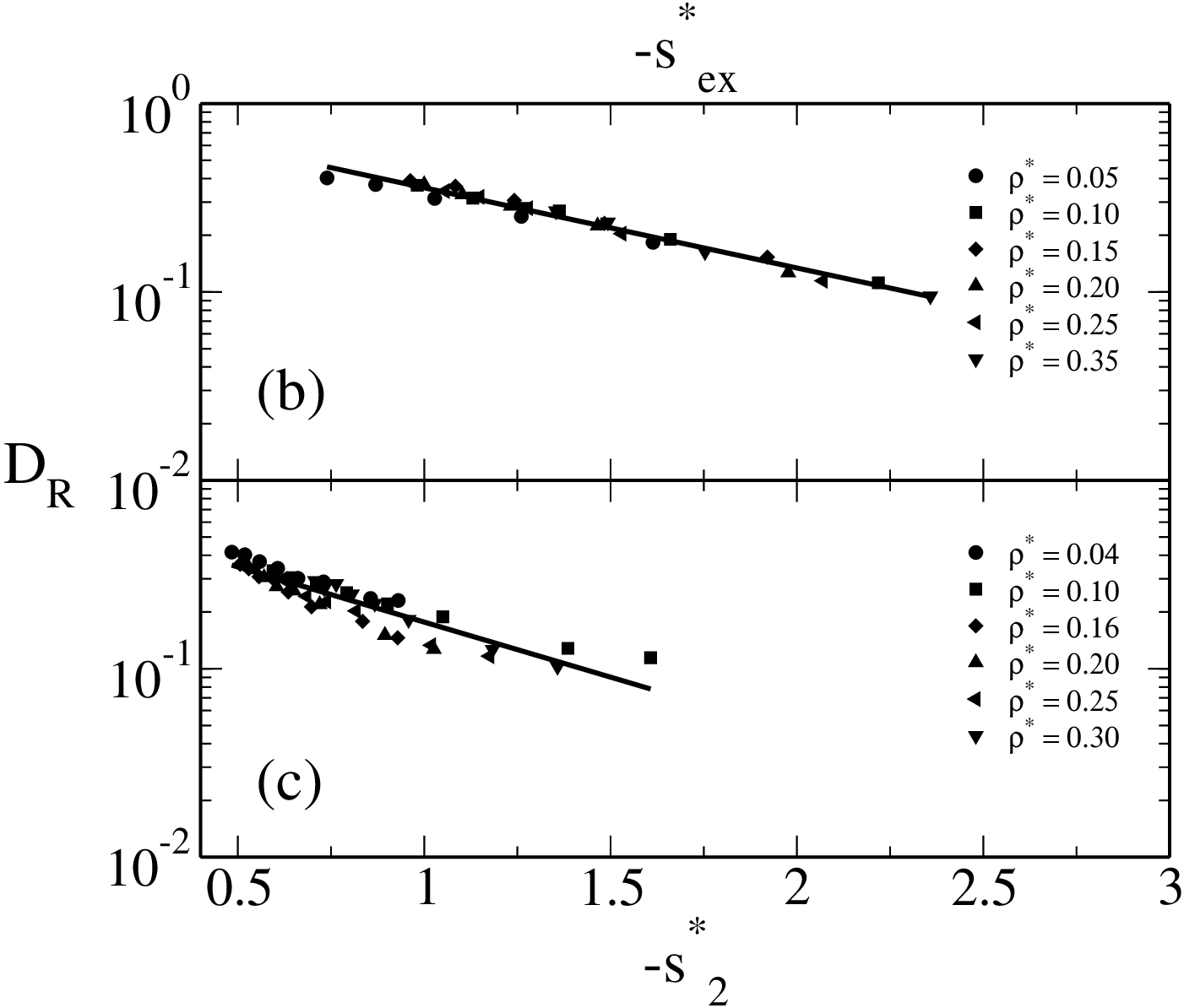}
\caption{ (a) Diffusion versus reduced density for fixed temperatures
$T^*=0.2,0.23,0.30,0.35,0.40,0.45,0.50,0.55$ from bottom to top;
(b) diffusion in Rosenfeld units versus the negative of the
$-s_{ex}^*$ and  (c) diffusion in Rosenfeld units
as a function of $-s_{2}^*$.}
\label{fig:D-DR}
\end{figure}

We now consider the scaling relationship between the
diffusivity and the excess entropy.  The diffusivity as a
function of density for different isotherms is shown
in Fig.~(\ref{fig:D-DR})(a). Clear maxima and minima in the 
$D (\rho )$ curves can be identified at low temperatures.
Fig.~(\ref{fig:D-DR})(b)-(c) show the scaling of the
reduced diffusivity, $D_R$, with the excess, $s_{ex}^*$,
and pair, $s_2^*$, entropy. Using the Rosenfeld macroscopic reduction 
parameters for the 
length  as $\rho^{-1/3}$ and the thermal velocity as $(k_BT/m)^{1/2}$, 
the dimensionless difusivity is defined as
\begin{equation}
D_R \equiv  D\frac{\rho^{1/3}}{(k_BT/m)^{1/2}}\; .
\label{eq:D*sexc}
\end{equation}
The scaling of the reduced diffusivity, $D_R$ with
with $s_{ex}^*$ is excellent with   $D_R=Ae^{Bs_{ex}^*}$ 
where $A=0.95$ and $B=0.98$. The scaling with the 
pair entropy, $s_2^*$, shows a weak isochore-dependence 
and the line of best fit is obtained with  $A=0.68$ and $B=1.35$.
The comparison between the 
Fig.~(\ref{fig:D-DR})(b) and Fig.~(\ref{fig:D-DR})(c) 
indicates that for  our anomalous fluid, the diffusivity scaling
with the pair correlation entropy is not as universal as
with the excess entropy. This is likely to be a consequence of the
presence of two, density-dependent length scales, in such water-like 
fluids when compared to simple liquids, such as the Lennard-Jones fluid.

\subsection{Instantaneous Normal Mode Analysis}

\begin{figure}[htbp]
   \centering
   \includegraphics[clip=true,scale=0.35]{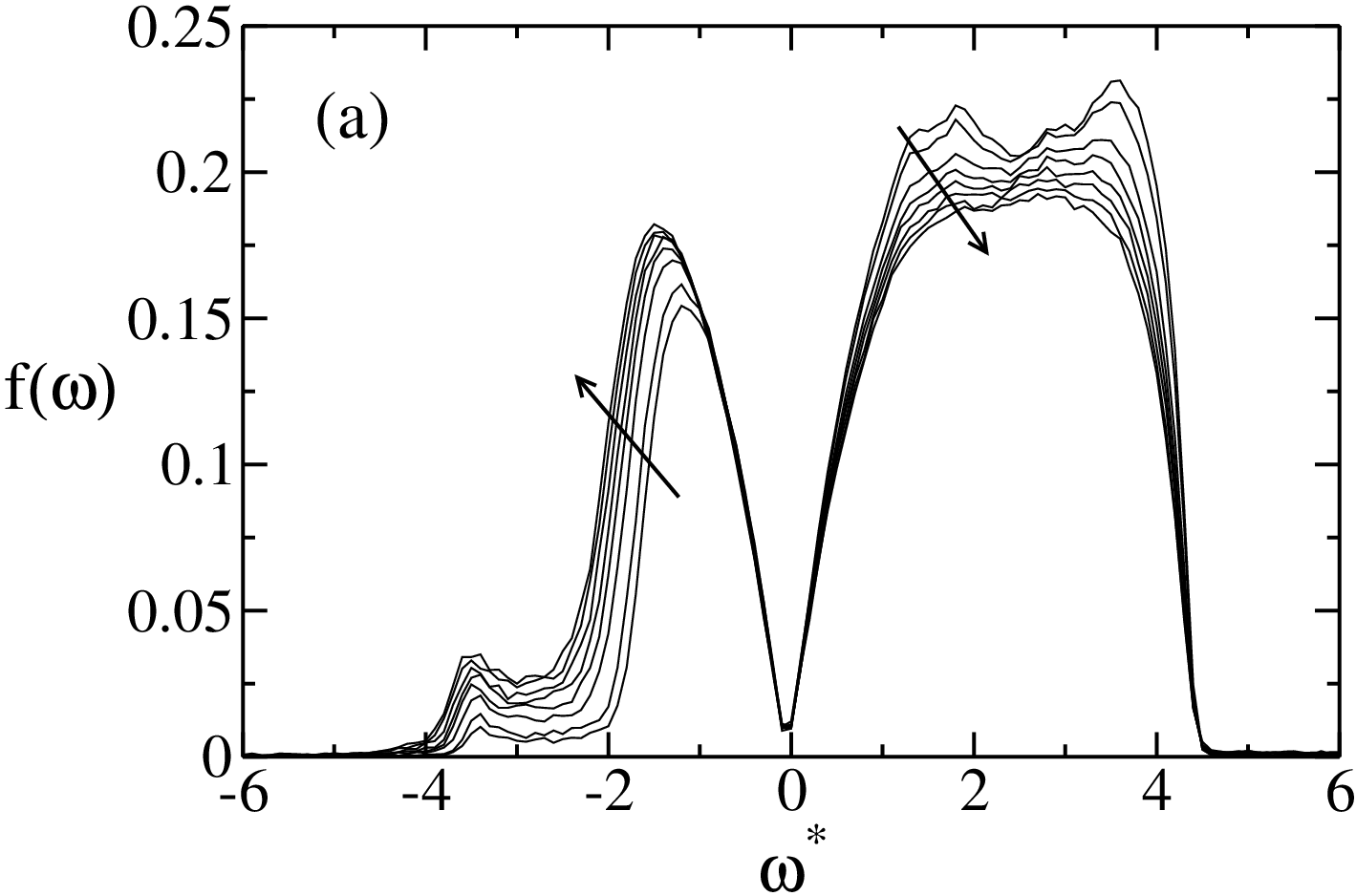}\includegraphics[clip=true,scale=0.35]{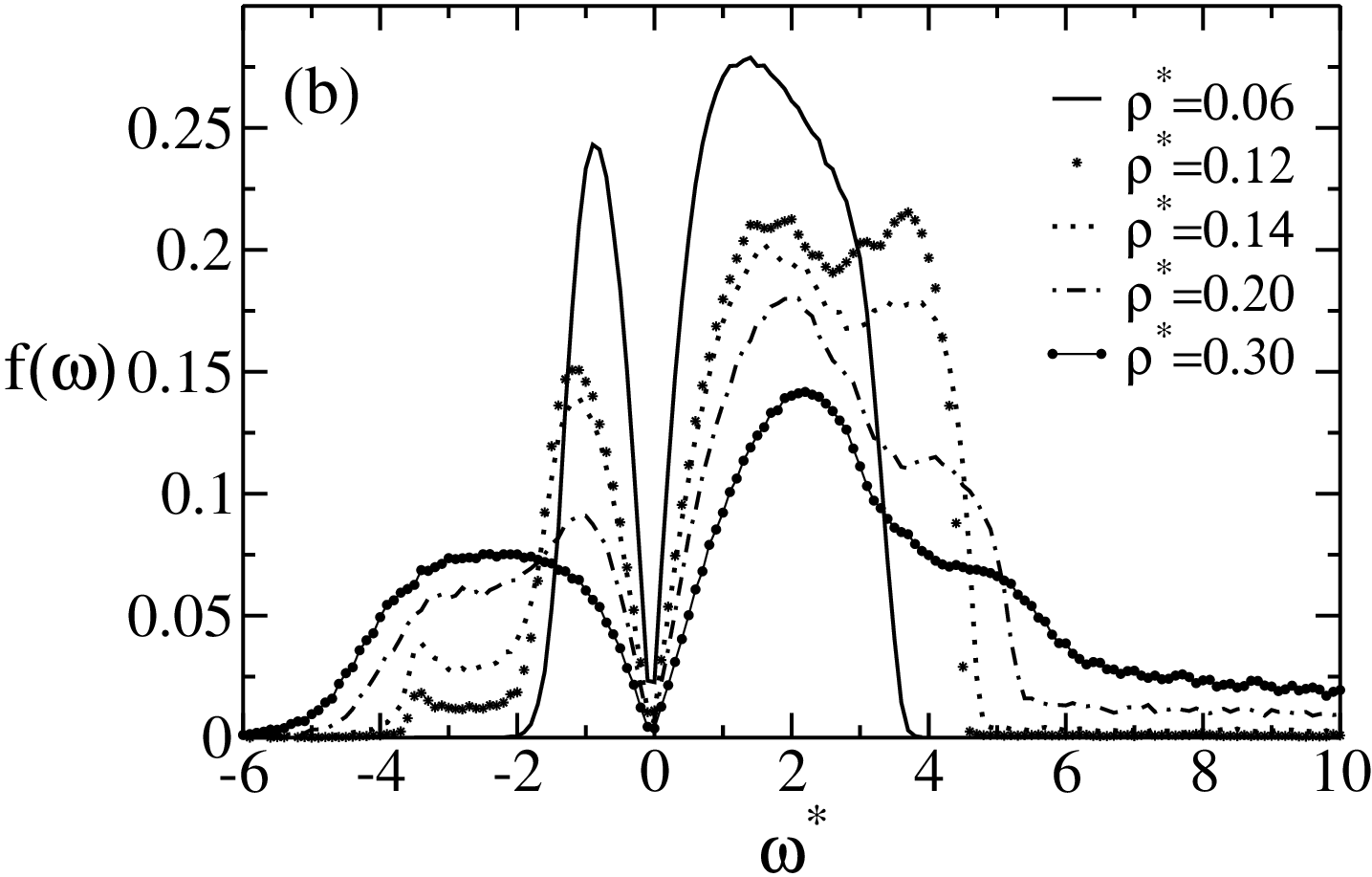} 
   \caption{(a) Instantaneous normal mode spectra as a function of frequency for
$T^*=$0.20,0.23,0.30,0.35,0.40,0.45,0.50,0.55   and 
$\rho^*=0.11$. The arrows  indicate the increase
of the temperature. (b) Instantaneous normal mode spectra as a function of frequency for
$T^*=0.20 $  and 
$\rho^*=$0.06,0.12,0.14,0.20,0.30.}
   \label{fig:rho-omega-0.11}
\end{figure}

Next, we present the results from our instantaneous
normal mode (INM) analysis of the simulation  for the
core-softened fluid. The INM spectra 
along the $\rho^*=0.11$ isochore for various temperatures
 is
illustrated in Fig.~\ref{fig:rho-omega-0.11}(a) while the INM spectra
along the $T^*=0.20$ isotherm for various densities is illustrated in
Fig.~\ref{fig:rho-omega-0.11}(b).
The shape of the INM spectra have the characteristic real and imaginary
branches. 
As in the case of  Lennard-Jones and Morse liquids,\cite{Sh01,Sh02} 
the negative modes
shrink in intensity and go to low frequencies as the temperature is 
decreased, while the peak of the real branch increases.
Unlike in the case of simple liquids, however, 
both the real and the imaginary branch have a pronounced bimodality
which can be clearly seen for the spectra along the $\rho^* =0.11$ isochore.
which must be connected
with two different length scales of the potential.
Fig.~\ref{fig:rho-omega-0.11}(b) shows that along the $T^*=0.20$ isotherm,
the bimodality in the imaginary branch is most pronounced within the anomalous regime and is
attenuated at both low and high densities. In contrast, the bimodality of the
real branch persists even at high densities. It would be interesting
to explore in future work if this bimodal frequency distribution
 results in multiple-time-scale behaviour 
analogous to that seen in hydrogen-bonded liquids, such as water and methanol.\cite{mcr06,scm08,akc10}

\begin{figure}[htbp]
   \centering
   \includegraphics[clip=true,scale=0.5]{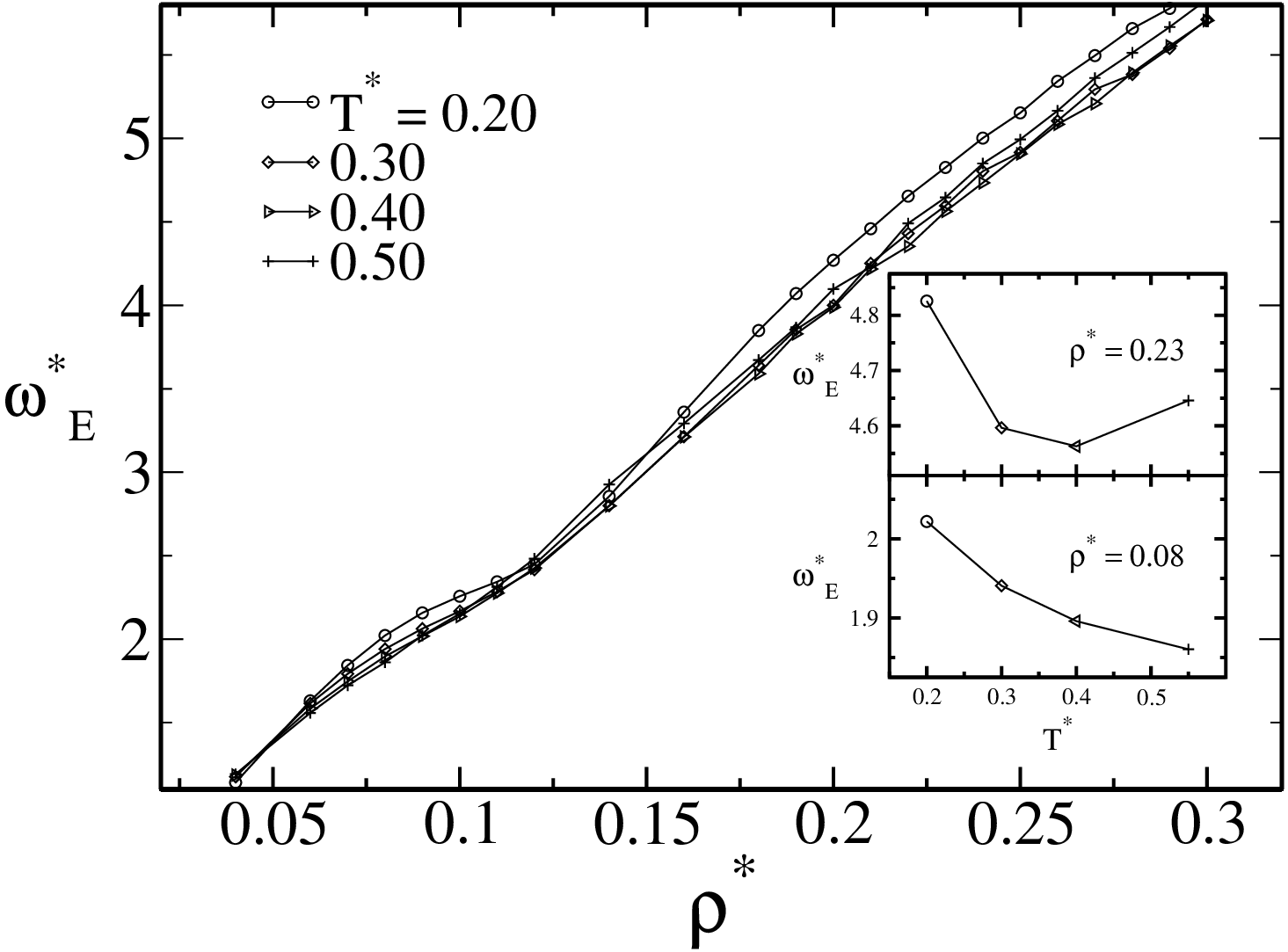} 
   \caption{Einstein Frequency versus density for fixed temperature. 
The insets show the dependence with temperature for two fixed
densities $\rho^*=0.08$ (bottom inset) and $\rho^*=0.23$ (top inset).}
   \label{fig:Wevsrho}
\end{figure}

The Einstein frequency is the second moment of the INM distribution,
as defined in equation Eq.~(\ref{eq:omegaE}), and 
represents an effective frequency
describing the very short-time, local dynamics of the particles.
Figure~\ref{fig:Wevsrho} illustrates the behavior 
of  Einstein frequency for the core-softened potential.
For a fixed temperature, increasing density results in 
increase of $\omega_E^*=\omega\sqrt{m\sigma^2/\epsilon}$, indicating stronger trapping of the
liquid particles in local cages. This is consistent with the behaviour
of simple liquids observed in earlier studies.\cite{Sh01,Sh02}
The $\omega_E^*$ values is 
virtually independent of  temperature for $\rho^*\approx 0.05$ and $\rho^*
\approx 0.125$. At low densities,  $\omega_E^*$ shows a small decrease with increasing temperature while at higher densities, there is a weak minimum in the $\omega_E^*$
at intermediate temperatures. The density  dependence of $\omega_E^*$ carries no significant
signatures of the diffusivity anomaly.

\begin{figure}[htbp]
   \centering
   \includegraphics[clip=true,scale=0.32]{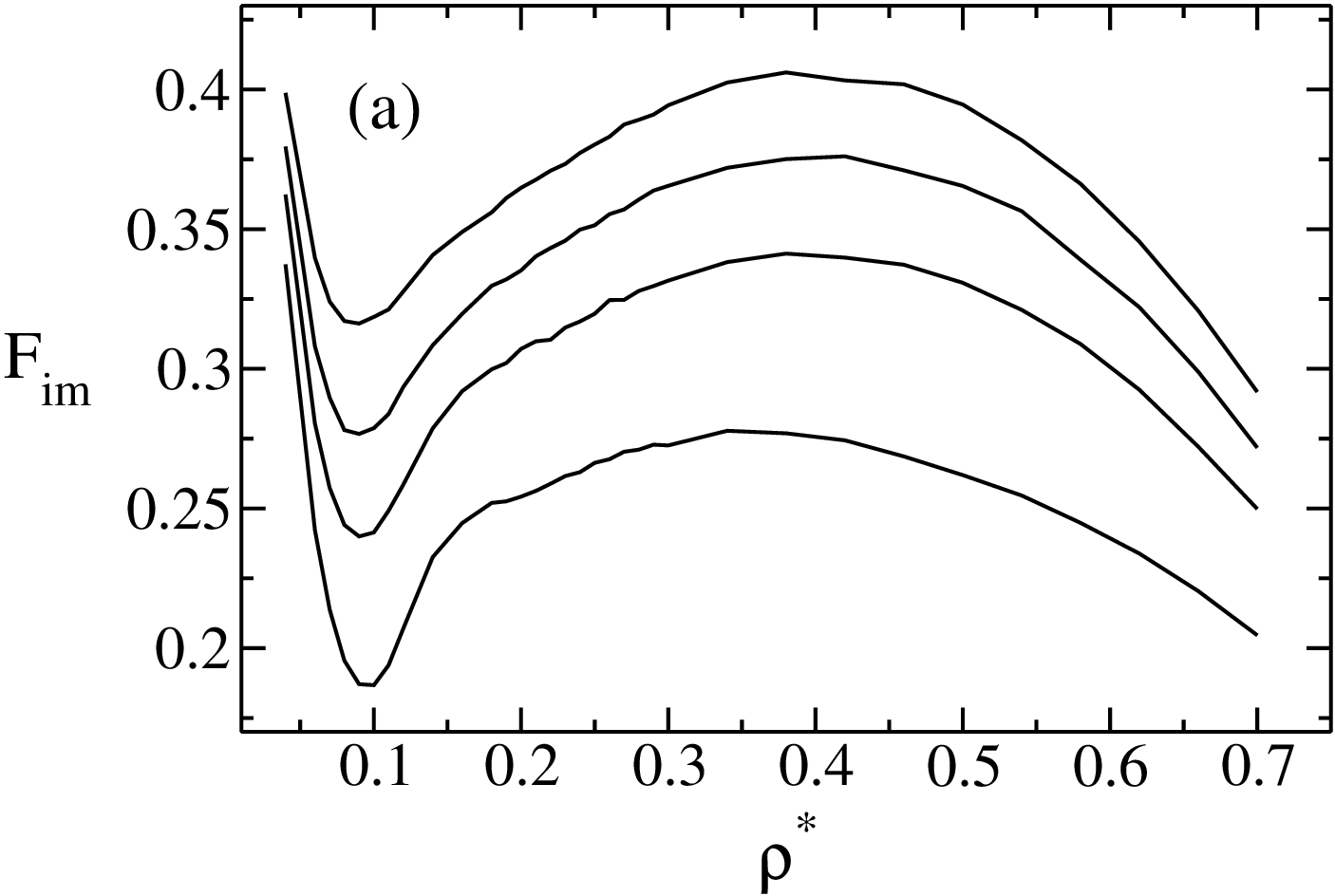} \includegraphics[clip=true,scale=0.32]{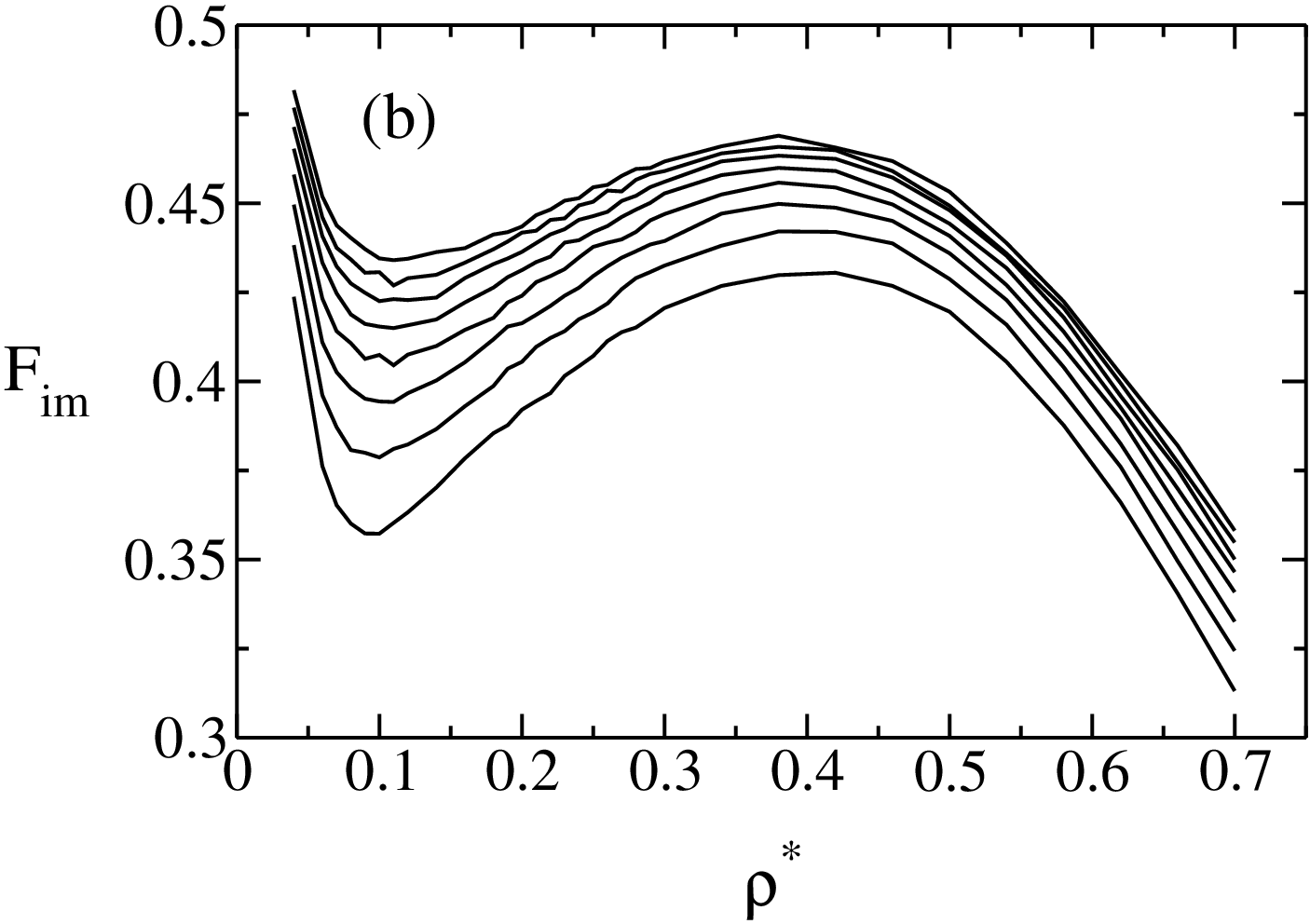}
   \caption{(a) Fraction of imaginary modes against density for  fixed 
temperatures $T^*=$0.20,0.30,0.40,0.55 from bottom to top. (b)
Fraction of imaginary modes against density for  fixed 
temperatures $T^*=$0.80,1.0,1.20,1.40,1.60.1.80,2.0,2.2  from bottom to top. 
  }
   \label{fig:Fivsrho}
\end{figure}

The fraction of imaginary modes,  $F_{im}$,  indicates
how much the system  samples regions with negative
curvature which is known to
be strongly correlated with the diffusivity. \cite{Se89} 
Fig.~\ref{fig:Fivsrho} shows density dependence 
fraction of imaginary modes, $F_{im}$, for our core-softened anomalous fluid. 
In simple liquids, $F_{im}$ decreases with density
and the graph $F_{im}$ vs. $\rho^*$ always exhibit
a negative slope.\cite{Sh01,Sh02} In contrast, for
 the core-softened potential studied here, $F_{im}$ shows very
pronounced non-monotonic behaviour   For very
low, $\rho<\rho_{Fmin}$, and very high, $\rho>\rho_{Fmax}$, densities, 
$F_{im}$ has
a negative slope, decreasing with
increasing density. For intermediate densities, $\rho_{Fmax}>\rho>\rho_{Fmax}$,
$F_{im}$ increases with density. The density, $\rho_{Fmin}^*\approx 0.1$, is 
almost temperature independent. The location of the maximum in
$F_{im}(\rho)$ curve is
 $\rho_{Fmax}^*\approx 0.4$ at low 
temperatures, but shifts to lower densities with increasing $T^*$.

\begin{figure}[htbp]
   \centering
   \includegraphics[clip=true,scale=0.5]{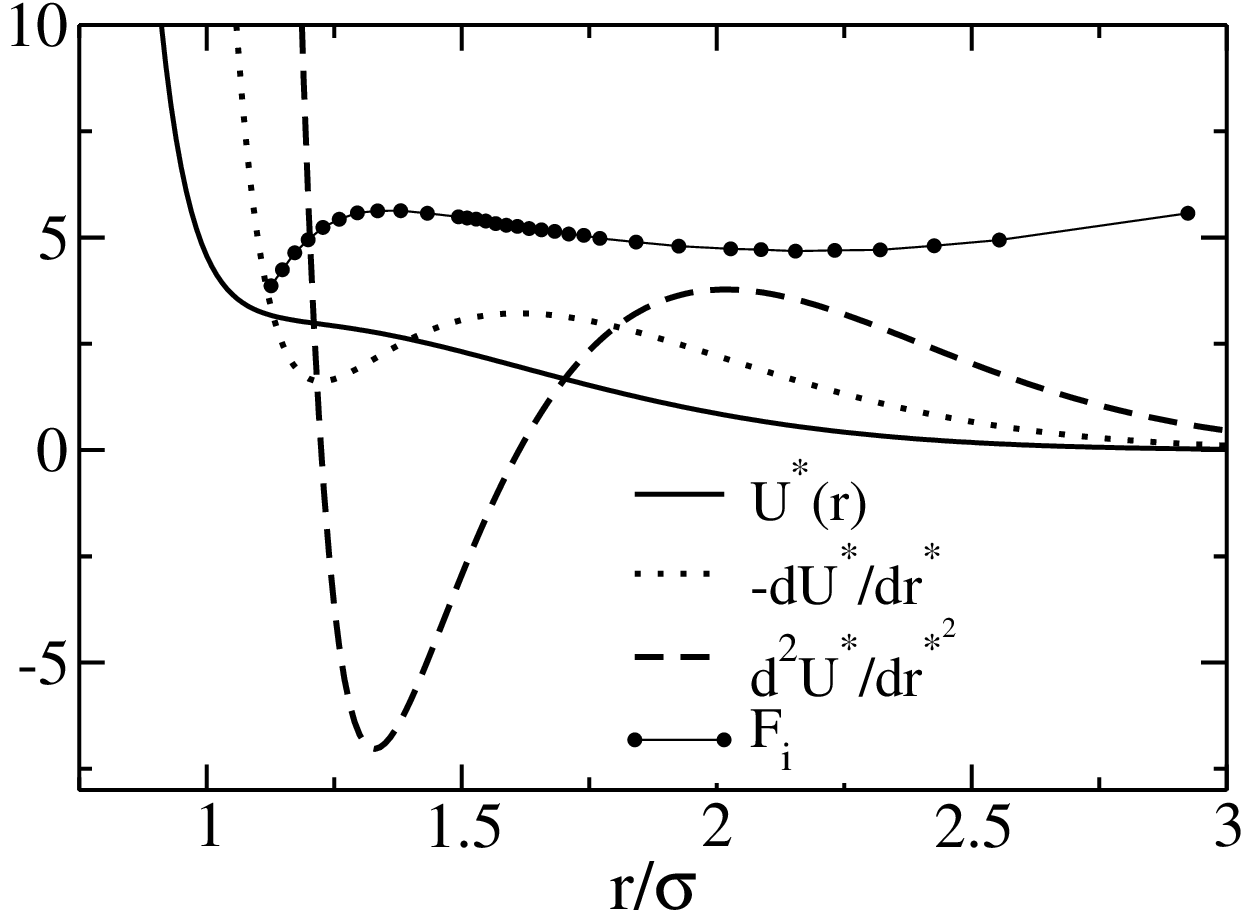} 
   \caption{Potential, force and  second derivative  of the potential
in units of $\epsilon$  and $F_{im}*15.0-1.0$ for $T^*=1.0$
versus reduced distance.}
   \label{fig:potential-force}
\end{figure}

A comparison of the behaviour of $F_{im}(\rho )$,
illustrated in Fig.~\ref{fig:Fivsrho}, and $D (\rho)$,
illustrated in Fig.~\ref{fig:D-DR}
 shows that the density of minimum $F_{im}$ coincides with
the density of minimum $D$. In contrast,  the density of maximum 
$F_{im}$ occurs 
at densities much higher than the density of maximum diffusivity.
Moreover, the region in which $F_{im}$  shows an anomalous increase
with compression persists to very high temperatures, well above the
temperature for onset of structurally anomalous behaviour.
In order to understand this behaviour, we compare the 
zeroth, first and second derivatives of the potential as
a function of pair separation with $F_{im}$ for
$T^*=1.0$ plotted
as a function of the mean interparticle separation, $\rho^{-1/3}$
in Fig.~(\ref{fig:potential-force}). It is immediately obvious that
the location of the minimum of $F_{im}$ coincides with
the location of the minimum of the second derivative. For densities lower
than this minimum, the second derivative increases and
the number of imaginary modes decreases. Clearly, this effect persists
in the high-temperature fluid where binary collisions dominate the dynamics
since  it reflects the curvature of the pair interaction.

\begin{figure}[htbp]
   \centering
   \includegraphics[clip=true,scale=0.5]{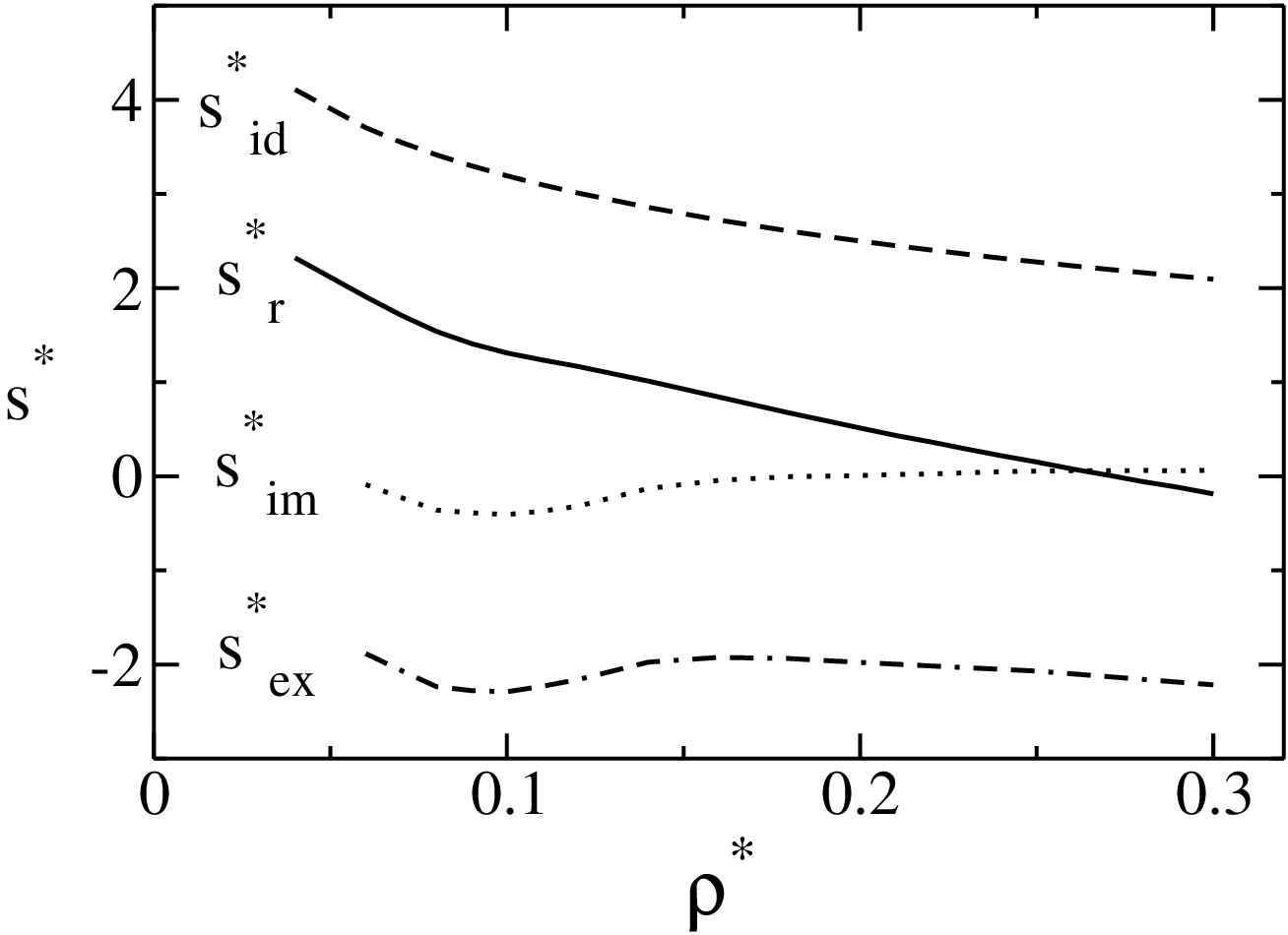} 
   \caption{ $s_{id} *$, $ s_{r}^*$, $s_{im}^*$ and $s_{ex}^*$  versus density for $T^*=0.2$.}
   \label{fig:s}
\end{figure}

\section{INM Spectra and Liquid-State Entropy}

In section III, we discuss the possibility of partitioning
the entropy of a liquid into contributions, $S_r$ and $S_{im}$,
associated with real and imaginary branches respectively of 
the INM spectrum. The $S_r$ contribution is directly derived from
the frequency distribution of the real branch while the $S_{im}$
contribution is given by $S_{im} =S-S_r$.
In order to get a better understanding of the role played
by each contribution to  the entropy, Fig.~(\ref{fig:s}) 
shows  the behavior with density for a fixed temperature of 
$s_{im}^*$, $s_r^*$, $s_{id}^*$ and $s_{ex}^*$. It is immediately 
evident that $s_r^*$ shows a very similar density dependence to
$s_{id}^*$, even though the numerical value  of $s_r^*$ is 
significantly lower. Other than a small inflection at $\rho^*\approx 0.1$,
the contribution of the real INM modes
to the entropy carries virtually no signature of the structural, entropy 
or diffusivity anomalies. The non-monotonic behaviour of $s_{ex}^*$
in the anomalous regime seems to be reflected only in $s_{im}^*$.
The strong resemblance between $s_{id}^*$ and $s_r^*$ in $T$- and $\rho$-dependence suggests that this term reflects the generic effects of thermal
fluctuations and spatial confinement on the entropy but in general it
is insensitive to the structural details associated with the
interplay between the two length scales in the anomalous regime.

\begin{figure}[htbp] 
   \centering
   \includegraphics[clip=true,scale=0.5]{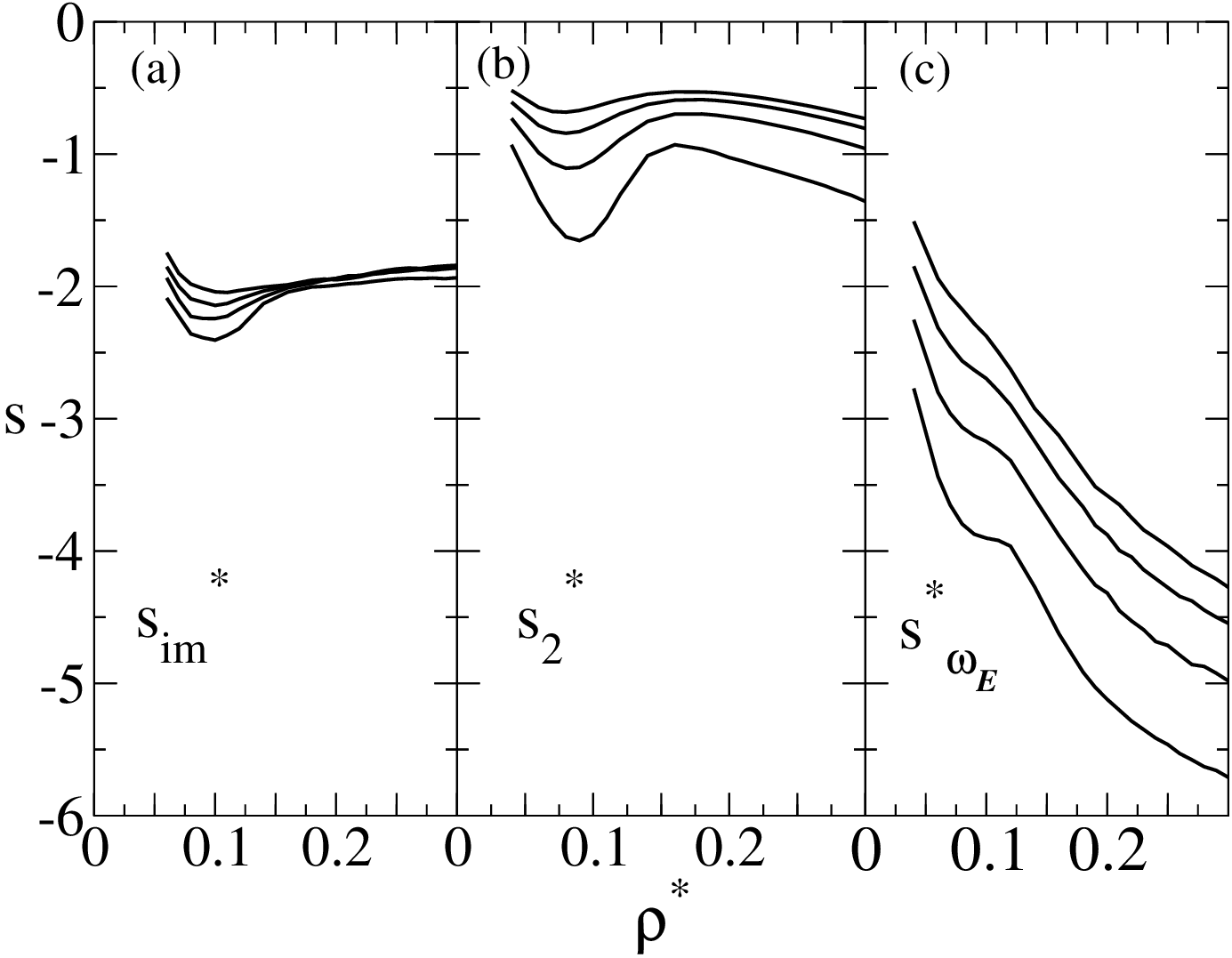} 
   \caption{(a) $s_{im}^*-2.0$ obtained from Eq.~(\ref{eq:simag}) with $s_{ex}^*$
obtained from ref. \cite{etm06}, 
   (b) $s_{2}^*$  and  (c) from the Einstein Frequency. See the text for more details.
   In all cases, the entropy is plotted against density for fixed temperatures
$T^* =$ 0.20, 0.30, 0.40 and 0.50  from bottom to top.}
   \label{fig:example}
\end{figure}

\begin{figure}[htbp]   
\centering   
\includegraphics[clip=true,scale=0.5]{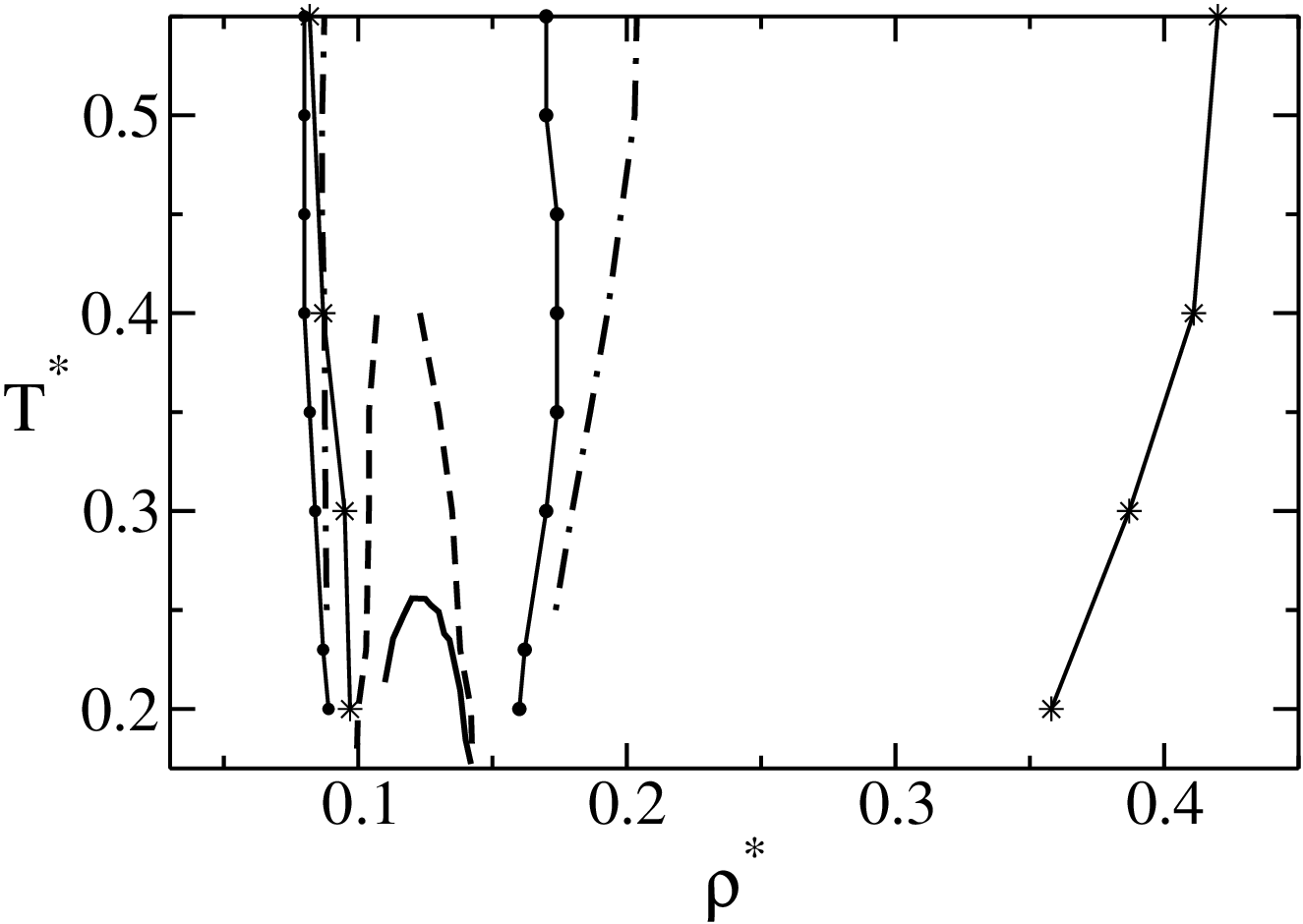}
\caption{Cascade of water-like anomalies in the density-temperature
plane. The solid line limits the region of 
density anomaly, the dashed line illustrates the region
of diffusion anomaly and the dot-dashed line 
shows the region of structural anomaly. The filled circles represent
the density of minimum and maximum $s_2$  and the stars represent
the region of minimum and maximum $F_{im}$.  }
\label{fig:extrema}
\end{figure}

Fig.~\ref{fig:example} compares, for various
isotherms,  the  density dependence 
of three different entropy estimators:  
(a) $s_{im}^*$, the imaginary mode contribution to the 
entropy, (in the graph, form the value of $s_{im}^*$ it
is subtracted 2), (b) $s_2^*$, the pair correlation entropy, and
(c) $s_{\omega_E}^*$, the entropy estimated using the
Einstein model for the liquid, Eq.~(\ref{eq:s-e}). 
As discussed in Section IV.A, the $s_2^*$ behaviour
is very similar in density and temperature dependence
to $s_{ex}^*$, indicating that the pair correlation contribution
to the entropy is sufficient to capture the essential
features of the anomalies. $s_{\omega_E}^*$,
derived from the Einstein model, is very similar to
$s_r^*$, presumably because of the use of the harmonic
oscillator representation of the entropy. The non-monotonic
behaviour of $\omega_E (\rho )$ results in small 
plateau at low temperatures. The overall decrease in
$S_{\omega_E}$ with density is much too steep and monotonic
compared with $s_{ex}^*$ and $s_2^*$. In contrast,
$s_{im}^*= s^*-s_r^*$, captures the behaviour of the
entropy within the anomalous region very successfully.
The relative displacement of $s_{im}^*(\rho)$ curves
for different isotherms is very small, consistent with
the earlier observation that the overall effect
of thermal and free volume effects is better 
captured by $s_r^*$.

As a summary of the insights obtained from
INM analysis and pair correlation entropy 
into the nested cascade of anomalies picture of water-like liquids,
we use the extrema in $F_{im}(\rho )$ and $s_2^*(\rho )$ curves
to define additional anomalous regimes within 
which $(\partial F_{im}/\partial\rho)_T >0$ and 
 $(\partial s_2^*/\partial\rho^*)_T >0$ respectively, as shown in
Fig.\ref{fig:extrema}.
The low density boundaries of anomalous regime in $t$,
$F_{im}$  and  $s_2^*$ almost coincide, reflecting the
onset of the steep repulsive wall. The high-density boundary of
the anomalous regime of  $F_{im}$ occurs at very high densities.
In contrast, the high density limit of the anomalous
regime in $s_2^*$ is very close to that defined by $t$,
and also by $s_{im}^*$ though the latter is not shown in the
Figure.
\section{Conclusions}

This paper explores the connection between
entropy, diffusivity and the potential energy landscape of
a core-softened fluid with water-like anomalies using
molecular dynamics simulations and instantaneous normal
mode analysis.

We demonstrate that the diffusivity and the excess
entropy  of a core-softened fluid with isotropic pair interactions
 obey Rosenfeld-type  excess entropy scaling of transport
properties. The use of macroscopic reduction parameters
for the diffusivity based on temperature and density
is particularly appropriate for fluids with multiple
length scales where defining an effective hard-sphere radius 
is inappropriate. We also show that the substituting 
the excess entropy by the pair correlation entropy leads
to a weak isochore dependence of the Rosenfeld-scaling parameters,
not seen in simple liquids but observed in other water-like 
liquids.\cite{acpre10}

The instantaneous normal mode spectra, including the 
Einstein frequency and the fraction of imaginary modes,
is computed over a wide range of temperatures and densities.
INM analysis is shown to provide unexpected insights into 
the dynamical consequences  of the interplay between length
scales characteristic of anomalous fluids that cannot be
obtained from an equilibrium transport property such as the
diffusivity.   

Both the real and imaginary branches of the
INM spectra exhibit bimodality that has so far not been
observed. As a function of density along an isotherm, the bimodality
in the real branch of the INM spectrum persists to very high
densities well beyond the structurally anomalous region. 
In contrast, the bimodality of the imaginary branch is much
more closely correlated with the region of the structural
anomaly. The bimodal character of the both branches of the INM
spectrum suggests that such core-softened fluids may show
multiple time-scale behaviour similar to that seen in hydrogen-bonded
systems.  

The Einstein frequency shows an essentially monotonic
dependence on density along an isotherm. The temperature
dependence of the Einstein frequency is weak and monotonically
decreasing with temperature at low densities and non-monotonic
at high temperatures. In contrast to the Einstein frequency, 
the fraction of imaginary frequencies shows very anomalous
behaviour in comparison to simple liquids, with an extended
density regime over which $F_{im}$ increases with increasing density.
While the low density boundary of this region coincides
with that of the structural anomaly, the high density boundary
occurs at very high densities well beyond the structurally 
anomalous regime. Previous INM studies of liquids have largely
connected the diffusivity  with the fraction of imaginary modes.
Our results show that information about diffusivity is
 largely contained in
the behaviour of the imaginary branch of the INM spectra,
but factors such as the bimodality of the frequency distribution
in this branch must be considered in addition to $F_{im}$. 

Given the validity of excess entropy scaling for the diffusivities,
we introduce INM-based estimators of the entropy, to connect the
energy landscape with liquid state thermodynamics and kinetics.
The conceptually simplest INM-based estimator is to treat the liquid
as a collection of 3-dimensional harmonic oscillators vibrating 
at a single frequency, corresponding to the Einstein frequency.
The Einstein model entropy shows a very steep decrease with density
along isotherms with a very weak signature of the onset of the
structurally anomalous regime. 

An alternative approach to developing an INM-based estimator of
the entropy that we
have explored is to assume that the total entropy of the
fluid can be written as a sum of contributions from $3NF_r$
harmonic modes and $3NF_{im}$ imaginary modes. The real
branch of the INM spectrum can be used to estimate the
harmonic contribution, $s_r^*$, exactly. The entropy contribution of the
imaginary branch, $s_{im}^*$, is then given by the difference of the
thermodynamic entropy, $s $, and the real branch contribution, $s_r^*$.
The temperature and density dependence 
of $s_r^*$  carries  virtually no signature of the liquid-state anomalies, and 
seems to reflect only the generic effects of thermal
fluctuations and spatial confinement on the entropy 
 In contrast, $s_{im}^*= s^* -s_r^*$, captures the behaviour of the
entropy within the anomalous region very successfully though 
the relative displacement of $s_{im}^*(\rho )$ curves
for different isotherms is too small. The extrema in $s_{im}^*$
define a region of anomalous entropy behaviour in the density-temperature
plane that is almost identical as the region within which $(\partial S_2
/\partial \rho)_T > 0$. 

The overall and somewhat unexpected outcome of our instantaneous
normal mode analysis of a core-softened water-like fluid 
is that the real and imaginary frequency branches show very different
sensitivities to the dynamical consequences of the interplay between
two length scales in the anomalous regime of the liquid.
Moreover, the entropy contribution from the imaginary frequency modes
of the INM spectrum 
reflects the anomalous behaviour of the excess entropy and diffusivity
characteristic
of water-like fluids, but the real frequency branch does not.

\subsection*{Acknowledgments}

This work is is supported by the Indo-Brazil Cooperation Program
in Science and Technology of the CNPq, Brazil and DST, India.This work is also
partially supported by CNPq, INCT-FCx.
The authors would like to thank Murari Singh for help with preparing
some of the figures.

\bibliographystyle{aip}
\bibliography{Biblioteca,inm,ccpub}

\begin{thebibliography}{10}

\bibitem{sw82}
F.~H. Stillinger and T.~A. Weber,
\newblock Phys. Rev. A {\bf 25}, 978 (1982).

\bibitem{ds01}
P.~G. Debenedetti and F.~H. Stillinger,
\newblock Nature (London) {\bf 410}, 259 (2001).

\bibitem{djw03}
D.~J. Wales,
\newblock {\em Energy Landscapes: With Applications to Clusters, Biomolecules
  and Glasses},
\newblock Cambridge University Press, Cambridge, 2003.

\bibitem{hm02}
J.-P. Hansen and I.~R. McDonald,
\newblock {\em Theory of Simple Liquids},
\newblock Academic Press, London, 2002.

\bibitem{Ja98}
E.~A. Jagla,
\newblock Phys. Rev. E {\bf 58}, 1478 (1998).

\bibitem{Wi06}
H.~M. Gibson and N.~B. Wilding,
\newblock Phys. Rev. E {\bf 73}, 061507 (2006).

\bibitem{Xu05}
L.~Xu, P.~Kumar, S.~V. Buldyrev, S.-H. Chen, P.~Poole, F.~Sciortino, and H.~E.
  Stanley,
\newblock Proc. Natl. Acad. Sci. U.S.A. {\bf 102}, 16558 (2005).

\bibitem{Ol07}
A.~B. de~Oliveira, M.~C. Barbosa, and P.~A. Netz,
\newblock Physica A {\bf 386}, 744 (2007).

\bibitem{Ol06a}
A.~B. de~Oliveira, P.~A. Netz, T.~Colla, and M.~C. Barbosa,
\newblock J. Chem. Phys. {\bf 124}, 084505 (2006).

\bibitem{Ol06b}
A.~B. de~Oliveira, P.~A. Netz, T.~Colla, and M.~C. Barbosa,
\newblock J. Chem. Phys. {\bf 125}, 124503 (2006).

\bibitem{Fo08}
D.~Y. Fomin, D.~Frenkel, N.~V. Gribova, and V.~N. Ryzhov,
\newblock J. Chem. Phys. {\bf 129}, 064512 (2008).

\bibitem{Eg08}
S.~A. Egorov,
\newblock J. Chem. Phys. {\bf 128}, 174503 (2008).

\bibitem{Mi06b}
J.~Mittal, J.~R. Errington, and T.~M. Truskett,
\newblock J. Chem. Phys. {\bf 125}, 076102 (2006).

\bibitem{Er01}
J.~R. Errington and P.~G. Debenedetti,
\newblock Nature (London) {\bf 409}, 318 (2001).

\bibitem{Ne01}
P.~A. Netz, F.~W. Starr, H.~E. Stanley, and M.~C. Barbosa,
\newblock J. Chem. Phys. {\bf 115}, 344 (2001).

\bibitem{Ne02b}
P.~A. Netz, F.~W. Starr, M.~C. Barbosa, and H.~E. Stanley,
\newblock J. Mol. Phys. {\bf 101}, 159 (2002).

\bibitem{Zh08}
S.~Zhou,
\newblock Phys. Rev. E {\bf 77}, 041110 (2008).

\bibitem{Ba09}
N.~M. Barraz, S.~E., and M.~C. Barbosa,
\newblock J. Chem. Phys {\bf 131}, 094504 (2009).

\bibitem{Ca03}
P.~Camp,
\newblock Phys. Rev. E {\bf 68}, 061506 (2003).

\bibitem{yr771}
Y.~Rosenfeld,
\newblock Phys. Rev. A {\bf 15}, 2545 (1977).

\bibitem{yr772}
Y.~Rosenfeld,
\newblock Chem. Phys. Lett. {\bf 48}, 467 (1977).

\bibitem{yr99}
Y.~Rosenfeld,
\newblock J. Phys-Cond. Matt. {\bf 11}, 5415 (1999).

\bibitem{md96}
M.~Dzugutov,
\newblock Nature (London) {\bf 381}, 137 (1996).

\bibitem{has00}
J.~J. Hoyt, M.~Asta, and B.~Sadigh,
\newblock Phys. Rev. Lett. {\bf 85}, 594 (2000).

\bibitem{sag01}
A.~Samanta, S.~M. Ali, and S.~K. Ghosh,
\newblock Phys. Rev. Lett. {\bf 87}, 245901 (2001).

\bibitem{sag05}
A.~Samanta, S.~M. Ali, and S.~K. Ghosh,
\newblock J. Chem. Phys. {\bf 123}, 084505 (2005).

\bibitem{scc06}
R.~Sharma, S.~N. Chakraborty, and C.~Chakravarty,
\newblock J. Chem. Phys. {\bf 125}, 204501 (2006).

\bibitem{ac09}
M.~Agarwal and C.~Chakravarty,
\newblock Phys. Rev. E. {\bf 79}, 030202 (2009).

\bibitem{agc09}
M.~Agarwal, A.~Ganguly, and C.~Chakravarty,
\newblock J. Chem. Phys. {\bf 113}, 15284 (2009).

\bibitem{etm06}
J.~R. Errington, T.~M. Truskett, and J.~Mittal,
\newblock J. Chem. Phys. {\bf 125}, 244502 (2006).

\bibitem{gpmc08}
T.~Goel, C.~N. Patra, T.~Mukherjee, and C.~Chakravarty,
\newblock J. Chem. Phys. {\bf 129}, 164904 (2008).

\bibitem{hsg52}
H.~S. Green,
\newblock {\em The Molecular Theory of Fluids},
\newblock North-Holland Pub. Co., 1952.

\bibitem{ng58}
R.~E. Nettleton and H.~S. Green,
\newblock J. Chem. Phys. {\bf 29}, 1365 (1958).

\bibitem{hjr71}
H.~J. Raveche,
\newblock J. Chem. Phys. {\bf 55}, 2242 (1971).

\bibitem{dcw87}
D.~C. Wallace,
\newblock J. Chem. Phys. {\bf 87}, 2282 (1987).

\bibitem{be89}
A.~Baranyai and D.~J. Evans,
\newblock Phys. Rev. A {\bf 40}, 3817 (1989).

\bibitem{sm96}
R.~M. Stratt and M.~Maroncelli,
\newblock J. Phys. Chem {\bf 100}, 12981 (1996).

\bibitem{rms95}
R.~M. Stratt,
\newblock Acc. Chem. Res. {\bf 28}, 201 (1995).

\bibitem{tk97}
T.~Keyes,
\newblock J. Phys. Chem. A {\bf 101}, 2921 (1997).

\bibitem{sc01a}
P.~Shah and C.~Chakravarty,
\newblock J. Chem. Phys. {\bf 115}, 8784 (2001).

\bibitem{sc021}
P.~Shah and C.~Chakravarty,
\newblock J. Chem. Phys. {\bf 116}, 10825 (2002).

\bibitem{sc022}
P.~Shah and C.~Chakravarty,
\newblock Phys. Rev. Lett. {\bf 88}, 255501 (2002).

\bibitem{cfsos94}
M.~Cho, G.~R. Fleming, S.~Saito, I.~Ohmine, and R.~M. Stratt,
\newblock J. Chem. Phys. {\bf 100}, 6672 (1994).

\bibitem{lssss00}
E.~L. Nave, A.~Scala, F.~W. Starr, H.~E. Stanley, and F.~Sciortino,
\newblock Phys. Rev. Lett. {\bf 84}, 4605 (2000).

\bibitem{rm97}
M.~C.~C. Ribeiro and P.~A. Madden,
\newblock J. Chem. Phys. {\bf 106}, 8616 (1997).

\bibitem{lks98}
W.-X. Li, T.~Keyes, and F.~Sciortino,
\newblock J. Chem. Phys. {\bf 108}, 252 (1998).

\bibitem{Mu05}
A.~Mudi, C.~Chakravarty, and R.~Ramaswamy,
\newblock J. Chem. Phys. {\bf 122}, 104507 (2005).

\bibitem{Th76}
H.~Thurn and J.~Ruska,
\newblock J. Non-Cryst. Solids {\bf 22}, 331 (1976).

\bibitem{LosAlamos}
Periodic table of the elements,
\newblock \url{http://periodic.lanl.gov/default.htm}, 2007.

\bibitem{Sa67}
G.~E. Sauer and L.~B. Borst,
\newblock Science {\bf 158}, 1567 (1967).

\bibitem{Ke83}
S.~J. Kennedy and J.~C. Wheeler,
\newblock J. Chem. Phys. {\bf 78}, 1523 (1983).

\bibitem{Ts91}
T.~Tsuchiya,
\newblock J. Phys. Soc. Jpn. {\bf 60}, 227 (1991).

\bibitem{An00}
C.~A. Angell, R.~D. Bressel, M.~Hemmatti, E.~J. Sare, and J.~C. Tucker,
\newblock Phys. Chem. Chem. Phys. {\bf 2}, 1559 (2000).

\bibitem{Ru06b}
R.~Sharma, S.~N. Chakraborty, and C.~Chakravarty,
\newblock J. Chem. Phys. {\bf 125}, 204501 (2006).

\bibitem{Sh02}
M.~S. Shell, P.~G. Debenedetti, and A.~Z. Panagiotopoulos,
\newblock Phys. Rev. E {\bf 66}, 011202 (2002).

\bibitem{Po97}
P.~H. Poole, M.~Hemmati, and C.~A. Angell,
\newblock Phys. Rev. Lett. {\bf 79}, 2281 (1997).

\bibitem{Ta02}
H.~Tanaka,
\newblock Phys. Rev. B {\bf 66}, 064202 (2002).

\bibitem{asc07}
M.~Agarwal, R.~Sharma, and C.~Chakravarty,
\newblock J. Chem. Phys. {\bf 127}, 164502 (2007).

\bibitem{ac07}
M.~Agarwal and C.~Chakravarty,
\newblock J. Phys. Chem. B {\bf 111}, 13294 (2007).

\bibitem{goddard}
S.-T. Lin, M.~Blanco, and W.~A. Goddard,
\newblock J. Chem. Phys. {\bf 119}, 11792 (2003).

\bibitem{Sh01}
P.~Shah and C.~Chakravarty,
\newblock J. of Chem. Phys. {\bf 115}, 8784 (2001).

\bibitem{mcr06}
A.~Mudi, C.~Chakravarty, and R.~Ramaswamy,
\newblock J. Chem. Phys. {\bf 122}, 104507 (2006),
\newblock Erratum: JCP, v124, p069902, 2006.

\bibitem{scm08}
R.~Sharma, C.~Chakravarty, and E.~Milotti,
\newblock J. Phys. Chem. B {\bf 112}, 9071 (2008).

\bibitem{akc10}
M.~Agarwal, H.~R. Kushwaha, and C.~Chakravarty,
\newblock J. Phys. Chem. B {\bf 114}, 651 (2010).

\bibitem{Se89}
G.~Seeley and T.~Keyes,
\newblock J. of Chem. Phys. {\bf 91}, 5581 (1989).

\bibitem{acpre10}
M.~Agarwal, M.~Singh, R.~Sharma, M.~P. Alam, and C.~Chakravarty,
\newblock (Preprint)  (2010).

\end{thebibliography}

\end{document}